\documentclass[fleqn,usenatbib]{mnras}

\usepackage{newtxtext,newtxmath}
\usepackage{graphicx}	
\usepackage{amsmath}	

\usepackage[T1]{fontenc}

\DeclareRobustCommand{\VAN}[3]{#2}
\let\VANthebibliography\thebibliography
\def\thebibliography{\DeclareRobustCommand{\VAN}[3]{##3}\VANthebibliography}

\def\subinrm#1{\sb{\mathrm{#1}}}
{\catcode`\_=13 \global\let_=\subinrm}
\mathcode`_="8000
\def\supinrm#1{\sp{\mathrm{#1}}}
{\catcode`\^=13 \global\let^=\supinrm}
\mathcode`^="8000

\title[The Signature of Refreshed Shocks from GRB030329]{The Signature of Refreshed Shocks in the of Afterglow of GRB030329}

\author[M. J. Moss et al.]{
Michael J. Moss,$^{1,2}$\thanks{E-mail: mikejmoss3@gmail.com (GWU)}
Robert Mochkovitch,$^{1}$
Fr\'ed\'eric Daigne,$^{1}$
Paz Beniamini,$^{3,4}$
and Sylvain Guiriec$^{1,5}$
\\
$^{1}$The Department of Physics, The George Washington University, 725 21st NW, Washington, DC 20052, USA\\
$^{2}$Sorbonne Universit\'e, CNRS, UMR 7095, Institut d'Astrophysique de Paris, 98 bis Arago, 75014 Paris, France\\
$^{3}$Department of Natural Sciences, The Open University of Israel, PO Box 808, Ra’anana 4353701, Israel\\
$^{4}$Astrophysics Research Center of the Open university (ARCO), The Open University of Israel, PO Box 808, Ra’anana 4353701, Israel\\
$^{5}$Astrophysics Science Division, NASA Goddard Space Flight Center, Greenbelt, MD 20771, USA
}

\date{Accepted XXX. Received YYY; in original form ZZZ}

\pubyear{2023}

\begin{document}
\label{firstpage}
\pagerange{\pageref{firstpage}--\pageref{lastpage}}
\maketitle

\begin{abstract}
GRB030329 displays one clear and, possibly, multiple less intense fast-rising ($\Delta t / t \sim 0.3$) jumps in its optical afterglow light curve. The decay rate of the optical light curve remains the same before and after the photon flux jumps. This may be the signature of energy injection into the forward and reverse shocked material at the front of the jet. In this study, we model the Gamma-Ray Burst (GRB) ejecta as a series of shells of material. We follow the dynamical evolution of the ejecta as it interacts with itself (i.e., internal shocks) and with the circumburst medium (i.e., external forward and reverse shocks), and we calculate the emission from each shock event assuming synchrotron emission. We confirm the viability of the model proposed by \citet{2003Natur.426..138G} in which the jumps in the optical afterglow light curve of GRB030329 are produced via refreshed shocks. The refreshed shocks may be the signatures of the collisions between earlier ejected material with an average Lorentz factor $\bar{\Gamma}\gtrsim 100$ and later ejected material with $\bar{\Gamma} \sim 10$ once the early material has decelerated due to interaction with the circumburst medium. We show that even if the late material is ejected with a spread of Lorentz factors, internal shocks naturally produce a narrow distribution of Lorentz factors ($\Delta\Gamma/\Gamma\lesssim0.1$), which is a necessary condition to produce the observed quick rise times of the jumps. These results imply a phase of internal shocks at some point in the dynamical evolution of the ejecta, which requires a low magnetization in the outflow.
\end{abstract}


\section{Introduction}  

Gamma-ray Bursts (GRBs) are bright flashes of gamma rays seen across our Universe, resulting from catastrophic events such as core-collapse supernova and compact binary mergers \citep{1998Natur.395..670G,2003Natur.423..847H,2013Natur.500..547T,2017PhRvL.119p1101A}. GRB030329 is a nearby GRB with detailed observations completed during both its prompt and afterglow emission phases. The afterglow displays a break at $\sim 0.5$ day which is likely to be the jet break, i.e., when the jet is decelerated to the point that the beaming factor of the emission is similar to the opening angle of the jet (i.e., $\Gamma\theta_j\sim1$), this results in the entire surface of the jet becoming visible to an observer \citep{1997ApJ...487L...1R,1999ApJ...525..737R}. The optical afterglow light curve shows several jumps occurring after the break, $\sim 1 - 6$ days after the initial trigger, all with rise times $\sim 0.3 - 0.8$ days (see Figure \ref{fig: comb mag light curves}). \citet{2003Natur.426..138G} discussed multiple possible mechanisms to create these jumps. Here, we concentrate on the ``refreshed shock'' mechanism, i.e., each jump is caused by slow moving material launched by the central engine eventually catching up with the decelerating material at the front of the jet \citep{1998ApJ...496L...1R,2000ApJ...535L..33S,2003Natur.426..138G}.

Using a simple 1-D model of the dynamical evolution of the ejecta, we confirm the viability of the refreshed shock model to explain the jumps observed in the optical afterglow of GRB030329. Furthermore, we show that the slow material responsible for the energy injection event must have a narrow distribution of Lorentz factors and that internal shocks naturally produce such a distribution. 

In Section \ref{sec: GRB030329 obs}, we outline the observations of GRB030329 and describe in more detail the interesting behavior witnessed in the afterglow. We describe our model in Section \ref{sec: sim codes}. In Section \ref{sec: analytical considerations}, we summarize the analytical considerations for the material responsible for the refreshed shocks. In section \ref{sec: afterglow jump sims}, we compare the features in GRB030329 to the jumps produced by our simulation and the implications of our findings. We conclude in Section \ref{sec: conclusion}.

\section{GRB030329 Observations} \label{sec: GRB030329 obs}

On March 29, 2003, GRB030329 was detected by HETE-II at $T_0$ = 11:37:14.67 UTC (41834.67 s UT) \citep{2003GCN..1997....1V}. The prompt emission displayed two pulses with similar fluences and a combined duration of $\sim50$ seconds. Soon after, a bright optical afterglow with magnitude R=12.6 was detected at $T_0$+1.5 hr. The source was followed up and observed by many facilities at wavelengths ranging from radio to X rays \citep{2003GCN..1985....1P,2003ApJ...597L.101T,2003GCN..1996....1M,2003GCN..2014....1B,2003GCN..2088....1H,2003GCN..2089....1K}. GRB030329 was a bright, nearby (z$\sim 0.1685$; \citealt{2003GCN..2020....1G}) GRB with a spectroscopically confirmed supernova \citep{2003ApJ...591L..17S}. The optical afterglow light curve decayed as a broken power-law with indices $\alpha_1 \approx -0.9$ and $\alpha_2 \approx -1.9$ with a break at $\lesssim0.5$ days (see Figure \ref{fig: GRB030329 light curves}; \citealt{2003Natur.426..138G}). 

\subsection{Jet Break and Angular Spreading}

The break in the optical light curve observed at $\lesssim0.5$ days is consistent with a break in the X-ray light curve \citep{2003A&A...409..983T}. The fact that the break is achromatic and that it is relatively smooth supports the assertion that this break is a jet break \citep{1999ApJ...519L..17S,2002MNRAS.332..945R}. 

A type Ic supernova was observed \citep{2003GCN..2107....1M,2003ApJ...599..394M}, which suggests that the progenitor was a Wolf-Rayet star and, therefore, that the surrounding medium has a stellar wind density profile (i.e., if the particle density goes as $\rho \propto r^{-k}$, where $r$ is the distance from the source, then $k=0$ describes a medium with a uniform density profile and $k=2$ describes a stellar wind density profile, assuming a constant mass-loss rate and wind velocity from the progenitor; \citealt{1992ApJ...391..246P,2007ARA&A..45..177C}). However, from radio observations, neither a uniform or a wind-like particle density profile are preferred for the surrounding circumburst medium \citep{2003Natur.426..154B,2005ApJ...634.1166V,2012ApJ...759....4M}.

The observed temporal indices before the jet break can reveal the density profile of the circumburst medium. For an external forward shock, it is expected that optical observations are within the frequency band $\nu_m < \nu < \nu_c$, where $\nu_m$ is the frequency associated with the minimum Lorentz factor of the accelerated electron population and $\nu_c$ is the characteristic synchrotron cooling frequency. In this frequency regime, if we assume that an electron population is accelerated to a distribution with a power index $p=2.2$, then the temporal decay index before a jet break will be $\alpha_1 = -0.9$ for a uniform medium and $\alpha_1 = - 1.4$ for a wind medium \citep{2002ApJ...571..779P}.

Looking at the optical afterglow light curve of GRB 030329, it appears that GRB 0303029 occurred in a uniform density medium. Furthermore, the jet break is a relatively sharp one, which also seems to support a uniform medium \citep{2003Natur.426..138G}. Assuming a uniform medium, the density is found to be $n\sim1$ cm$^{-3}$ from multi-wavelength afterglow modeling \citep{2003Natur.426..154B,2005ApJ...634.1166V,2012ApJ...759....4M}.

The Konus instrument on board the Wind satellite measured the fluence in the $15 - 5000$ keV band to be $1.6 \times 10^{-4}$ erg cm$^{-2}$ \citep{2003GCN..2026....1G}. Using the observed redshift $z = 0.1685$ \citep{2003GCN..2020....1G}, the isotropic gamma-ray energy can be estimated to be $E_{\gamma,iso}\sim10^{52}$ erg \citep{2003Natur.426..138G}. Using the estimated values for $E_{\gamma,iso}$ and the external medium density $n$, the observed jet break time $t_j$ and redshift $z$, and assuming a fiducial value of the gamma-ray efficiency of $\eta_\gamma = 0.15$ \citep{2015MNRAS.454.1073B}, we can estimate the opening angle of the jet for a uniform density medium \citep{2001ApJ...562L..55F}, 

\begin{align}
	\theta_j &= 0.076 \left(\frac{t_j}{1 \text{ day}}\right)^{3/8} \left(\frac{1+z}{2}\right)^{-3/8} \left(\frac{E_{\gamma,iso}}{10^{52} \text{ erg}}\right)^{-1/8} \nonumber \\
	&\hspace*{1cm} \times \left(\frac{\eta_{\gamma}}{0.2}\right)^{1/8} \left(\frac{n}{0.1 \text{ cm}^{-3}}\right)^{1/8} \sim 0.07 \text{ rad}. \label{eq: theta_j}
\end{align}

After the jet break, for either a uniform or stellar-wind medium density profile, it is expected that $\alpha_2 = -p$ in the case of relativistic lateral expansion at the sound speed in the medium, i.e., $c_s = c / \sqrt(3)$, or $\alpha_2 = -3p/4$ for a jet with no lateral expansion \citep{2002ApJ...571..779P,2007RMxAC..27..140G}. In the case of GRB 030329, it is found that $p \sim 2.2$ \citep{2005ApJ...634.1166V,2012ApJ...759....4M}, making the expected post-jet-break index either $\alpha_2 = -2.2$ or $-1.65$, respectively. The observed value of the post-jet-break slope, $\alpha_{2,obs} \sim -1.9$, may indicate that the lateral expansion of GRB 030329 is somewhere in between these two extremes. 
Furthermore, numerical simulations suggest the lateral spreading is low in the case of GRB030329 \citep{2003Natur.426..138G}. For these reasons, we assume lateral expansion is not significant for GRB030329 at the time of the jumps.

\subsection{Jumps in the Afterglow}

\begin{figure}
	\centering
	\includegraphics[width=0.48\textwidth]{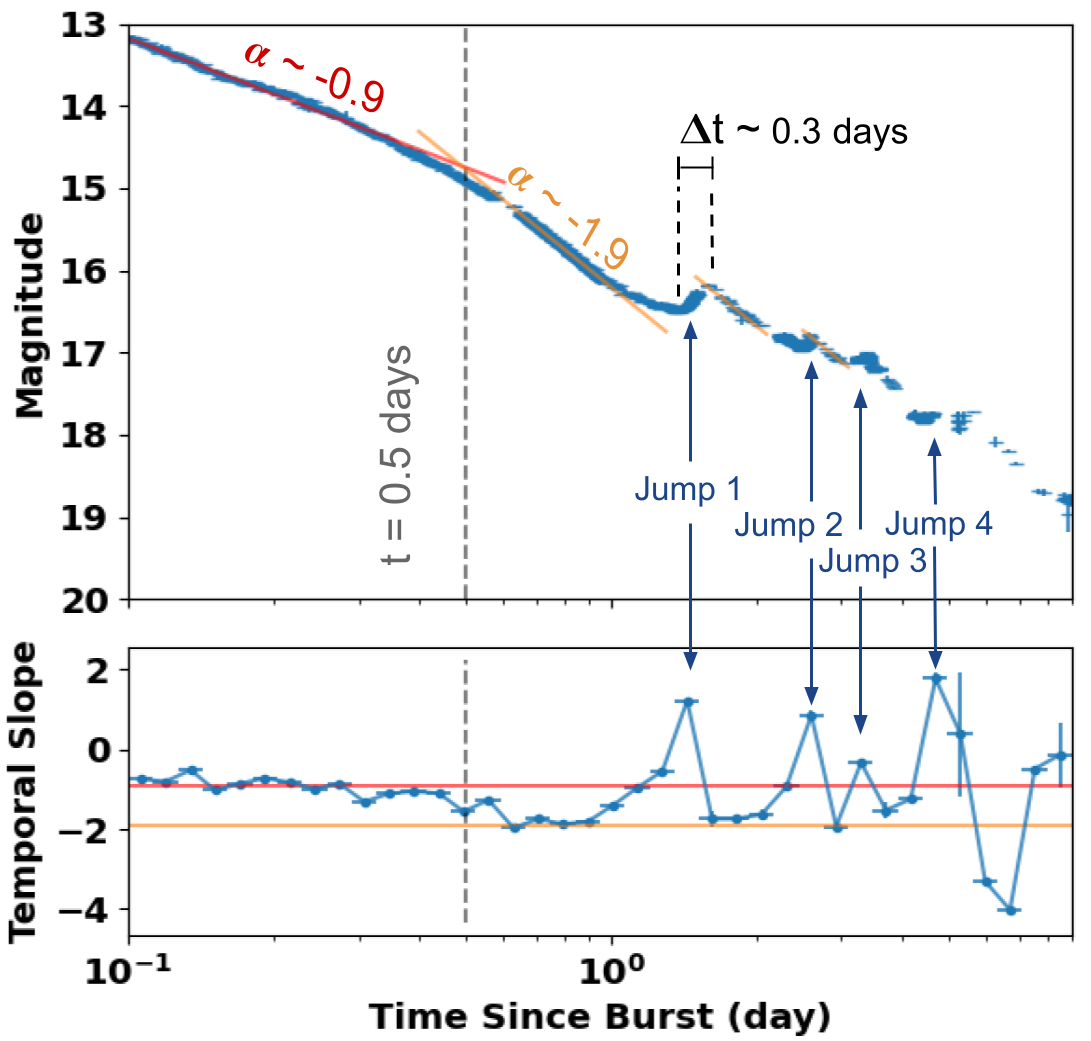}
	\caption{$(Top)$ The R-band afterglow light curve of GRB030329 (blue) created from aggregated data of multiple observatories (includes data from \citet{2003Natur.423..843U,2003AstL...29..573B,2003ApJ...599..394M,2004AJ....127..252B,2004ApJ...606..381L}). The gray dashed line indicates $t=0.5$ days, the jet-break time \citep{2003Natur.426..138G}. The red and orange lines are the approximate power laws of the light curve before and after the jet break, respectively. After each jump, the light curve returns to the same temporal index found after the jet break (i.e., $\alpha \sim -1.9$). $(Bottom)$ Displayed is the evolution of the temporal slope assuming a power law, $F \propto t^{\alpha}$.}
	\label{fig: GRB030329 light curves}
\end{figure}

Following the jet break, the optical light curve displays several jumps occurring between 1 and 6 days, all with rise times of $\sim 0.3 - 0.8$ days (see Figure \ref{fig: GRB030329 light curves}). Before and after each jump, the temporal decay index of the afterglow remains similar, but the amplitude of the emission is increased, hinting that the jumps are episodes of energy injection. In a decelerating blastwave model, the typical dynamical timescale in the ejecta is approximately equal to the time of observation, i.e., $\Delta t \sim t\sim R/2c\Gamma^2$, where $R$ is the radius at which the emission occurs and $\Gamma$ is the bulk Lorentz factor of the material. However, the jumps display short rise times, i.e., $\Delta t / t < 0.3$, which is difficult to satisfy. The first of these jumps occurs at $\sim 1.3$ days and has a rise time of $\Delta t\sim0.3$ days, and results in an increase in the observed flux by a factor of $\sim$2. The remaining jumps increase the flux by a factor of $\sim1.3 - 1.4$ \citep{2003Natur.426..138G}.

There are various mechanisms that may be responsible for producing these jumps: (i) sharp increases in the external medium density, (ii) non-uniformity in the energy distribution per unit solid angle of the jet, (iii) a multi-component jet (i.e., a laterally structured jet), and (iv) energy injection from material launched by the central engine. 

The first possibility is that the circumburst medium density increases with radius in multiple discrete steps. Wind termination shocks can produce sharp density increases in the density profile of the circumburst medium and could be observed as rises in the optical afterglow light curve. For standard parameters, flux rises due to interaction with stellar wind profiles will occur at a few days \citep{2006MNRAS.367..186E}. If the wind is episodic, several sharp density increases may be present and could possibly be observed as jumps similar to those observed for GRB030329. However, \citet{2007MNRAS.380.1744N}, show that for a wind-termination shock density medium, both the increased travel time of high-latitude photons as compared to those emitted along the line of sight and the dynamics of outflow material being crossed by a reverse shock smooth out the variability of the emitted light curve, such that the light curve would be much smoother than the sharply rising jumps of GRB 030329.

A second possibility is the so-called ``patchy shell'' model, where the energy per unit solid angle injected in the GRB outflow is not constant \citep{1996ApJ...473..998F,1997ApJ...485..270S,2004ApJ...602L..97N,2005ApJ...631..429I}. However, if we interpret the break in the light curve at $\lesssim 0.5 $ days as the jet break, then the whole emitting surface of the ejecta would be visible at the time of the jumps and the non-uniformity of the energy per unit solid angle should be negligible. 

The third option assumes that a narrow jet with opening angle $\theta_{n,j}$ and bulk Lorentz factor $\Gamma_n \simeq 200$ is launched surrounded by a wider jet with opening angle $\theta_{w,j} > \theta_{n,j}$ and bulk Lorentz factor $\Gamma_w \simeq 15$ \citep{1993ApJ...418..386L,1998ApJ...496..311P,2000ApJ...538L.129F,2003Natur.426..154B,2003ApJ...595L..33S,2003ApJ...594L..23V}. In some cases, the wider jet can become visible during the afterglow and appear as an episode of re-brightening \citep{2005ApJ...626..966P}. Indeed, modeling the radio data of GRB030329 requires breaks at $t\sim10$ and $\sim50$ days \citep{2003GCN..2014....1B,2005ApJ...634.1166V,2012ApJ...759....4M}, leading to a proposed scenario where multiple hard-edged jets were launched with concentrically increasing opening angles and decreasing bulk Lorentz factors. In order to possibly produce the multiple jumps observed in the optical afterglow of GRB030329, the lateral structure of the jet would need to involve several components. This model struggles to reproduce the sharp rise time observed for the jumps of GRB030329 \citep{2005ApJ...626..966P}.

The final option considered is the ``refreshed shocks'' model, which was most supported by \citet{2003Natur.426..138G}. In this model, slow moving material ($\Gamma \sim 10$) is launched by the central engine some time after the material responsible for the main prompt emission ($\Gamma \gtrsim 100$). The late ejecta will eventually catch up and collide with early ejecta after it decelerates due to interactions with the circumburst medium, at which time the collisions will be witnessed as a series of jumps in the light curve, similar to those of GRB030329 \citep{2003Natur.426..138G,2020MNRAS.495.2979A}. We explore the refreshed shock model here.

\section{Simulating GRB Dynamics and Emission} \label{sec: sim codes}

\begin{figure}
	\centering
	\includegraphics[width=0.46\textwidth]{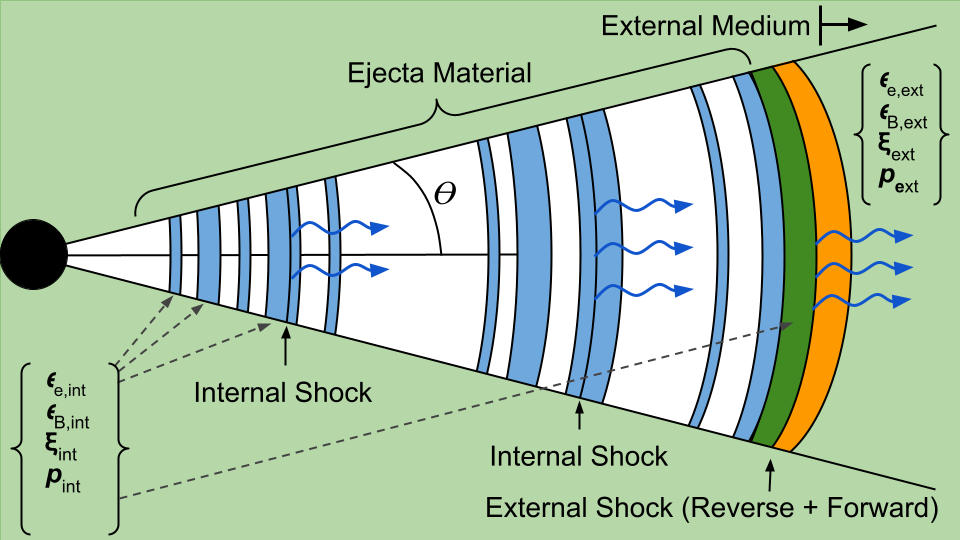}
	\caption{A schematic of the ballistic model we use to describe GRB outflow. GRB ejecta are represented as a series of shells launched with varying masses and Lorentz factors (in blue). Due to their varying speeds, the shells collide via internal shocks. The front most shell sweeps up material from the circumburst medium and decelerates, representing material being crossed by an external forward shock. Ejecta material that catches up to and collides with the front of the jet represents material being crossed by an external reverse shock. Light curves are generated as a summation of elementary emission contributions from each shock event.}
	\label{fig: schematic}
\end{figure}

In this work, we represent the relativistic outflow of the jet using a simple 1-D model where a series of thin shells are launched with varying Lorentz factors with an average Lorentz factor $\bar{\Gamma} \sim 150$ (see Figure \ref{fig: schematic}; \citealt{1997ApJ...490...92K,1998MNRAS.296..275D}). The mass of each shell is proportional to the inverse of its Lorentz factor, i.e., $1/\Gamma$, so that the isotropic equivalent energy injection rate of the outflow, $\dot{E}_{iso}$, is held constant. The ejection times of the shells are separated by a time $\delta t_e$ over a total wind duration $t_w$. When a shell with Lorentz factor $\Gamma_2$ and mass $m_2$ catches up and collides with a slower shell in front of it with mass $m_1$ (i.e., $\Gamma_2 > \Gamma_1$), the shells merge into a single shell with mass $m_r = m_1 + m_2$ and dissipate an energy $e_{diss}$ (i.e., $e_{diss} \ll m_1 \sim m_2$). After complete redistribution of momentum and energy between the two colliding shells, the Lorentz factor of the resulting shell is defined as $\Gamma_{r}$. The time and radius of collision (source frame) between two shells launched with a separation $\delta t_e$ in time between them can be estimated as,

\begin{align}
	t_{coll} &\approx \delta t_e \frac{(2\Gamma_1^2 - 1)}{(\Gamma_2^2 - \Gamma_1^2)}\Gamma_2^2 \label{eq: t coll is} \\
	R_{coll} &= c\beta_2t_{coll} \approx c \delta t_e \frac{(2\Gamma_1^2-1)(2\Gamma_2^2-1)}{2(\Gamma_2^2 - \Gamma_1^2)} \label{eq: r coll is}
\end{align}

where $\Gamma_2 > \Gamma_1$. The arrival time of emission generated by the shocks is calculated relative to a signal which would have traveled at the speed of light from the central engine to the observer, 

\begin{align}
	t_a = t_{coll} - R_{coll}/c \label{eq: t arriv}
\end{align}

We represent external forward and reverse shocks as two separate, but adjacent, regions at the front of the outflow that share the same bulk Lorentz factor $\Gamma_{r,ES}$ and location $R_{FS} = R_{RS}$. We model the forward shock as a set of discretized shock events that occur every time the outer-most shell in the ejecta sweeps up a mass $m_{sweep} = f \frac{M_{FS}}{\Gamma_{FS}}$, where $f\sim0.01$ sets the discretization size of the swept up material, $\Gamma_{FS}$ is the bulk Lorentz factor of the forward shock, and $M_{FS}$ is the mass of the external medium material crossed by the forward shock at the time of the previous forward shock event \citep{2007MNRAS.381..732G}. A reverse shock occurs as ejecta material catches up and collides with ejecta material previously decelerated at the shock discontinuity between the outflow and the external medium.

We assume a fraction $\epsilon_e$ of the dissipated energy is transferred to the synchrotron electrons and self-consistently calculate $\alpha_{synch}$, the fraction of electrons emitting via synchrotron radiation (the remaining electrons emit via inverse-Compton emission at higher energies, which is not considered in this work; \citealt{1998MNRAS.296..275D}). This makes the output energy of each shock event $E = e_{diss} \epsilon_e \alpha_{synch}$ \citep{1998MNRAS.296..275D,2007MNRAS.381..732G}. The emission of the burst is composed of a summation of elementary emission contributions, each with a bolometric luminosity profile, $l(t)$, which includes the curvature effect of the emitting shell, 

\begin{align} \label{eq: lum prof}
	l(t) = \frac{2E}{\Delta t_{obs}[ 1 + (t - t_{obs})/\Delta t_{obs} ]^3}
\end{align}

for $t_{obs} < t < t_{obs} + (1 - \cos\theta_j) R_{sh} / c$, where $\theta_j$ is the opening angle of the jet assumed to be seen on axis and $\Delta t_{obs} = R_{sh}/(2c\Gamma_r^2)$ is the spread of arrival times due to photons being emitted at different latitudes at the same radius \citep{1999ApJ...513..679G,1999ApJ...523..187W,2007MNRAS.381..732G}. We assume that the momentum and energy of two colliding shells are immediately redistributed across the combined material of the merged shells. Finding an accurate redistribution time requires detailed hydrodynamic simulations beyond the scope of this work \citep{2011MNRAS.415..279V,2018MNRAS.474.2813L,2020MNRAS.495.2979A}.

The local rest-frame density of material in the outflow is largely unknown and is the most uncertain quantity of our model. We estimate the comoving density of the ejecta material, $\rho_{ej}$, and circumburst material, $\rho_{cb}$, as

\begin{align}
	\rho_{ej} &\approx \frac{\dot{E}_{kin}}{4\pi R_{sh}^2 \Gamma_{ej}^2 c^3}\\
	\rho_{cb} &\approx 4\rho_0\Gamma_{r,ES}\left(\frac{R_*}{R_{FS}}\right)^k
\end{align}

where $R_*$ is the radius to the stellar surface, $\rho_0$ is the density of the circumburst material at the surface of the star, and $\dot{E}_{kin}$ is the isotropic equivalent kinetic energy injection rate is defined. The estimated density of the ejecta material is likely an underestimate since the material is compressed when shocked.

For a uniform medium, $\rho_{0,uni} = m_pn_0$ (where $n_0$ is the particle density of the medium) and for a wind medium $\rho_{0,wind} = A_{*} \dot{M}_{w} / 4 \pi v_{\infty} R_*^2$ (where $\dot{M}_{w} = 10^{-5} M_{\odot}$ yr$^{-1}$ is the mass loss rate of the stellar wind, $v_{\infty} = 10^8$ cm s$^{-1}$ is the asymptotic velocity of the wind, and we assume $R_* = 10^5$ cm, which are all typical values for Wolf-Rayet stars; \citealt{1986ARA&A..24..329C,1997ApJ...485L...5W,1997MNRAS.288L..51W,1999ApJ...524..262M,2003ApJ...595..935M}). The isotropic equivalent kinetic energy injection rate is defined as $\dot{E}_{kin} = \dot{E}_{iso}/(1+\sigma)$, where $\sigma$ is the large-scale magnetization of the material in the jet and is assumed to be small enough to allow for shocks to form (i.e., $\sigma < 0.1$).

The equipartition magnetic field density in the emitting material $B_{eq}$ can be approximated as

\begin{align}
	B_{eq,ej} &\simeq (8\pi \epsilon_B \rho_{ej} \epsilon_*)^{1/2} \\
	B_{eq,cb} &\simeq c\Gamma_{FS}(32\pi \epsilon_B \rho_{cb} )^{1/2} 
\end{align}

where $\epsilon_B$ is the fraction of the total energy contained in the magnetic field, $\epsilon_* \approx e_{diss}/(2 m \Gamma_r)$ is the average energy dissipated for each proton in the shock event, and $m$ is the mass of the material involved in the collision \citep{1998MNRAS.296..275D,1999ApJ...520..641S}. 

All emission from shocked material is assumed to be synchrotron radiation. In order to determine if the emission is in fast- or slow- cooling synchrotron regime, we calculate $\nu_c$ (the characteristic synchrotron cooling frequency) and $\nu_m$ (the frequency associated with the typical Lorentz factor of the electron population) for each shock event.

\begin{align}
	\nu_c &= \frac{(18\pi m_e q_e c)}{B_{eq}^3 \Delta t_{obs}^2 (1+Y)^2} \label{eq: nu_c}\\
	\nu_m &= \frac{1}{2\pi} \left(\frac{(p-2)}{(p-1)}\frac{\epsilon_e}{\xi}\right)^2 \frac{q_e m_p^2}{m_e^3 c^5} \epsilon_* B_{eq} \label{eq: nu_m}
\end{align}

where $m_e$ and $q_e$ are the electron mass and charge, respectively, $\xi$ is the fraction of electrons accelerated to a power law when the material is crossed by a shock front, and $Y$ relates the emitted power from synchrotron to synchrotron self-Compton \citep{1998ApJ...497L..17S,1999ApJ...523..177W,2009ApJ...703..675N}.

A fraction $\xi$ of electrons in the material crossed by the shock front are accelerated to a power law and cool via synchrotron radiation. Due to its degeneracy with other afterglow microphysical parameters, it is commonly assumed that the fraction of accelerated electrons is close to unity when considering the forward shock \citep{2005ApJ...627..861E}, but by using peaks in radio afterglow data, it has been found that $\xi_e$ can take a range of values, e.g. $10^{-1} \lesssim \xi \lesssim 1$ \citep{2017MNRAS.472.3161B,2023MNRAS.518.1522D}. Furthermore, there is evidence that the microphysical parameters of internal and external jet material should not be the same and, in fact, can be quite different. The need for a low $\xi$ for material internal to the jet is motivated by various synchrotron models for GRB prompt emission \citep{1998MNRAS.296..275D,2007NewA...12..630R,2009A&A...498..677B,2013ApJ...769...69B}. We assume the fraction of electrons accelerated by the reverse shock to be $\xi_{ej} = 10^{-2}$, since they are internal to the jet, while leaving the fraction of electrons accelerated by the forward shock as $\xi_{cb} = 1$. Equations $\ref{eq: nu_c}$ and \ref{eq: nu_m} indicate that $\nu_c$ does not depend on $\xi$ while $\nu_m$ strongly depends on $\xi$. This means that if the reverse shock is in the synchrotron slow-cooling regime (i.e., $\nu_m < \nu_c$), then by decreasing $\xi$ the low energy break will occur at higher frequencies and, for cases when $\xi$ is sufficiently low, the reverse shock may be in the synchrotron fast-cooling regime, where it becomes more radiatively efficient (i.e., $\nu_m > \nu_c$; see Figure \ref{fig: spectrum at 1 day}).

Similarly, we assign microphysical parameters value separately when considering material internal to the jet (i.e., material involved in internal and reverse shocks) and external material (i.e., material involved in for forward shocks; see Table \ref{tab: param values}). When considering the equations above, the relevant internal or external value of each microphysical parameter are used.

In order to model an ejecta that produces refreshed shocks, we suppose that the central source first ejects material with $\Gamma_{early} \geq 100$, which is responsible for the prompt emission, and, sometime later (e.g., $\sim 5 - 10$ seconds) eject slower material with $\Gamma_{late} \sim 10$ (see schematic in Figure \ref{fig: ref-shock-schem} and an example Lorentz factor profile in Figure \ref{fig: simulated lorentz dist}). As the outflow propagates, the early ejecta will sweep up circumburst material and will eventually decelerate until it is slow enough that the late ejecta can catch up and collide with the front of the jet, resulting in energy injection, i.e., refreshed shocks.

\section{Analytical Considerations for Refreshed Shocks} \label{sec: analytical considerations}
\subsection{Time of the Jump}

We can estimate the time of collision between the early and late material, i.e., the time of a jump, by using a toy model where the ejecta is composed of only two shells, a faster shell (with $\Gamma_f$) and a slower shell (with $\Gamma_s$) are launched at the same radius and at the same time. Assuming both shells expand spherically, the mass swept up by the fast shell at a radius $R$ can be expressed as 

\begin{align}
	M_{swept} &= \int_0^{R} 4\pi r^2 \rho(r) dr.
\end{align}

The faster shell travels with a constant Lorentz factor $\Gamma_f$ until it reaches the deceleration radius, $R_d$, i.e., the radius at which it has swept up a mass equal to $M_{swept} = M_f / \Gamma_f$ where $M_f \approx E/\bar{\Gamma} c^2$ is the mass of the fast shell, $E$ is injected energy, and $\bar{\Gamma}$ is the average Lorentz factor of the rapid part of the outflow. At the deceleration radius, the mass of material that has been swept up can be expressed by

\begin{align}
	M_{swept} &= 4\pi \int_0^{R_d} \rho_* \left(\frac{r}{R_*}\right)^{-k}r^{2} dr\\
	\frac{M_f}{\Gamma_f}&= \frac{4\pi \rho_0 R_*^k}{(3-k)} R_d^{3-k}. \label{eq: M_swept}
\end{align}
 
Rearranging for $R_d$, we can write 

\begin{align} \label{eq: decel rad}
	R_d &= \left(\frac{(3-k)E}{4\pi\rho_0 R_*^k\Gamma_f^2 c^2}\right)^{1/(3-k)}.
\end{align}

After reaching the deceleration radius, the fast material will follow the Blandford-Mckee solution. Following the prescription of \citet{2017A&A...605A..60B}, we can describe the radius and propagation time (source frame) of the fast shell as

\begin{align}
	t_{f} &= \int_0^R \frac{dr}{2c\Gamma^2} = \int_0^{R_d} \frac{dr}{2c\Gamma_f^2} + \int_{R_d}^R \frac{r^{2\epsilon} dr}{2c\Gamma_f^2 R_d^{2\epsilon}}\\
	&= \frac{t_d}{(2\epsilon + 1)} \left(2\epsilon + \left(\frac{R}{R_d}\right)^{2\epsilon+1}\right) \label{eq: t fast}
\end{align}

where $t_d = R_d / (2c\Gamma_f^2)$ is the deceleration time and $\epsilon = (3-k)/2$, such that, $\epsilon = 3/2$ for a uniform density medium and $\epsilon = 1/2$ for a wind density medium. 

The fast shell pulls ahead of the slow shell and will sweep up circumburst material, meaning the slow shell is ``protected'' by the fast shell so it will not sweep up material from the external medium nor decelerate. We can therefore simply describe the relation between the radius and time of the slow material as

\begin{align} \label{eq: t slow}
	t_{s} = \frac{R}{2c\Gamma_s^2}.
\end{align}

The collision will occur at a time $t_{coll}$ when the radii of the two shells are the same. We find this by substituting Equation \ref{eq: t slow} into Equation \ref{eq: t fast},

\begin{align} \label{eq: t coll}
	t_{coll} &= \frac{t_d}{(2\epsilon+1)}\left[2\epsilon + \left(\frac{2c\Gamma_s^2t_{coll}}{R_d}\right)^{2\epsilon+1}\right]\\
	&= \frac{t_d}{(2\epsilon+1)}\left[2\epsilon + \left(\frac{\Gamma_s^2t_{coll}}{\Gamma_f^2 t_d}\right)^{2\epsilon+1}\right].
\end{align}

In the limit where $t_{coll} \gg t_d$ (in the case of GRB 030329, $t_{coll} \approx 10^5 \text{ sec } \gg t_d \approx 30 \text{ sec }$), we obtain 

\begin{align} \label{eq: t coll simple}
	t_{coll} \approx t_d (2\epsilon + 1)^{1/2\epsilon} \left(\frac{\Gamma_f}{\Gamma_s}\right)^{2+1/\epsilon}.
\end{align}

Using Equation \ref{eq: decel rad} and the expression for $t_d$, Equation \ref{eq: t coll simple} can be written as

\begin{align}
	t_{coll, uni} &= \left(\frac{3E_f}{8 \pi \rho_{0,uni} c^5 \Gamma_s^8}\right)^{1/3}\\ 
	&\approx 10^5 \left(\frac{E_{f,53}}{n_0}\right)^{1/3} \Gamma_{s,1}^{-8/3} \text{ s} \label{eq: t coll intuit const} \\
	t_{coll, wind} &= \left(\frac{3E_f}{4 \pi \rho_{0,wind} c^3 \Gamma_s^4}\right)\\ 
	&\approx 10^5 \left(\frac{E_{f,53}}{A_*}\right) \Gamma_{s,1}^{-4} \text{ s} \label{eq: t coll intuit wind}
\end{align}

for a uniform medium and a stellar-wind medium, respectively, where $E_{f,53}$ is the isotropic equivalent energy of the fast material, $E_f = M_f\Gamma_f c^2$, in units of $10^{53}$ erg, and $\Gamma_{s,1}$ is the Lorentz factor of the slow material in units of $10$ \citep{2017A&A...605A..60B}. The estimated values are contingent on the assumed gamma-ray efficiency. We see that the time of the collision is strongly dependent on the Lorentz factor of the slow material, placing a tight constraint on the average Lorentz factor of the slow material.

\subsection[Estimation of Gamma_s]{Estimation of $\Gamma_s$} \label{sec: estimate gamma_s}

Using Equation \ref{eq: t coll intuit const}, \ref{eq: t coll intuit wind}, and the observed time of the first jump in the afterglow light curve of GRB030329 (i.e., $t_{jump}\sim 1$ day $\sim 10^5$ seconds, see Figure \ref{fig: GRB030329 light curves}), we can estimate the Lorentz factor of the slow material. Taking the measured values of the injected isotropic energy, $E_{iso,obs} = E_{f} \sim 10^{52}$ erg and reasonable external density values for a uniform medium (e.g., $n_0 = 1$ cm$^{-3}$) and a wind medium (e.g., $A_* = 1$), we find


\begin{equation}
\Gamma_s \gtrsim
    \begin{cases}
        8.5 & \text{(uniform medium)} \\
        6.4 & \text{(wind medium)}
    \end{cases}
\end{equation}

for a uniform medium and stellar-wind medium, respectively. 

\subsection{Rise Times of the Jumps} \label{sec: rise times}

When the late ejecta catches up to and collides with the material at the front of the jet, it will be crossed by the external reverse shock, adding energy to the entire external shock system. This energy injection will appear as a jump in the light curve as the radiation material transitions to a new Blandford-Mckee solution corresponding to an increase in total energy. 

The duration of this transition (i.e., $\Delta t_{jump}$, the rise time of the jumps) is determined primarily by the reverse shock crossing time, i.e., the time it takes the reverse shock to cross the incoming ejecta. Which is determined by the difference in Lorentz factors across the slow ejecta $\Delta \Gamma_s$. 

The duration of this transition (i.e., $\Delta t_{jump}$, the rise time of the jumps) is primarily determined by the difference in Lorentz factors across the slow ejecta $\Delta \Gamma_s$ since the rise time is the signature of the reverse shock crossing time, i.e., the time it takes the reverse shock to cross the incoming ejecta. 

\subsubsection{Dispersion of Lorentz Factors in the Slow Material} \label{sec: gamma disp}

Equation \ref{eq: t coll simple} gives the time of collision between shells of material with a single Lorentz factor. More realistically, the material will not have a uniform Lorentz factor but a distribution of Lorentz factors with some spread $\Delta\Gamma_s$. Taking the derivative of Equation \ref{eq: t coll simple}, we can approximately relate the duration of the jump with the spread of the Lorentz factors in the slow material:

\begin{align}
	\frac{\Delta t_{coll}}{t_{coll}} = \left(2+\frac{1}{\epsilon}\right) \frac{\Delta\Gamma_s}{\Gamma_s} \label{eq: gamma spread}
\end{align}
where $\Gamma_s$ is now the average Lorentz factor of the slower ejecta after it has undergone internal shocks. The first optical jump in GRB030329 was observed with $\Delta t / t \sim 0.3$ (see Figure \ref{fig: GRB030329 light curves}) which implies $\Delta\Gamma_s/\Gamma_s \sim 0.1$ for a uniform density medium (or $\Delta\Gamma_s/\Gamma_s \sim 0.075$ for a wind-like medium). This narrow spread of Lorentz factors required of the slow material may seem ad hoc and unrealistic, but it can naturally be obtained via the internal shock mechanism. 

Detailed hydrodynamic simulations show that internal collisions between jet material creates plateaus in the Lorentz profile of the jet \citep{2000A&A...358.1157D,2009A&A...498..677B,2020MNRAS.495.2979A}. During internal shocks, faster material catches up and collides with slower material. The total energy and momentum are then redistributed across the combined material of the merged shells. This will cause plateaus to form in the Lorentz factor profile of the ejecta material (see Figure \ref{fig: simulated lorentz dist}). Internal shocks will continue until the profile of Lorentz factors is monotonically increasing radially in the outflow. 

When the initially fast material begins to decelerate, a reverse shock propagates back into the outflow. The shocked material will begin to cool and emit radiation, the strength of which is determined by the mass of the incoming material and the amount that the material is decelerated. The rise time of the emission is determined by the energy redistribution timescale which can be approximated as the reverse shock crossing timescale in the observer frame \citep{1995ApJ...455L.143S}. In principle, this may affect the observed rise time of the emission found using the simplified estimate given by Equation \ref{eq: gamma spread}. However, since the reverse shock is relativistic, $\Delta t_{coll}/t_{coll} \approx 2 (\Delta \Gamma_s / \Gamma_s) (\Gamma_f/\Gamma_s)^2 \approx \Delta \Gamma_s / \Gamma_s$, where the Lorentz factors are defined at the time of collision. This suggests that the estimate given by Equation \ref{eq: gamma spread} is valid up to a factor of order unity.

Another important timescale to consider arises due to the geometric opening angle of the jet. However, the angular timescale determines the decay time of the emission, not the rise time \citep{2016MNRAS.459.3635B}. Photons emitted from relativistically moving jetted material will display a spread of arrival times due to photons emitted at different latitudes at the same radius and is approximated as $\Delta t \approx t \approx R/2c\Gamma^2$ \citep{1996ApJ...473..998F,2011MNRAS.415..279V}. After a jet break, the spread in arrival times should be approximated using the opening angle of the jet at the time of emission, i.e., $\Delta t \approx R\theta_j^2/2c$. In the context of refreshed shocks, the spread in arrival times will be determined by the geometry of the slow ejecta, not the fast ejecta, thus decoupling $\Delta t$ from $t$. The slow ejecta propagates in the wake of the fast material and so will not encounter any external medium. We can therefore assume that the opening angle of the slow material, $\theta_s$, will not change, i.e., $\theta_s = \theta_j$. 


\section{Late Afterglow Energy Injection due to Early Prompt Internal Shocks} \label{sec: afterglow jump sims}

\begin{figure}
	\centering
	\includegraphics[width=0.45\textwidth]{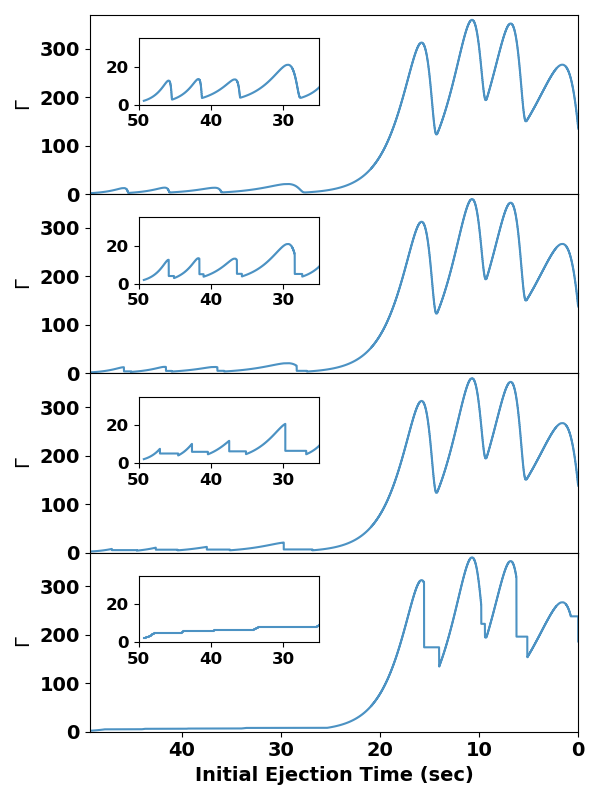}
	\caption{Stacked snapshots of the Lorentz factor profile of the ejecta at multiple times (earliest to latest snapshots from top to bottom, respectively). A zoom-in of the late ejecta material is displayed within the plot inset. Plateaus are seen to form in the Lorentz factor profile as internal shocks occur within the outflow.}
	\label{fig: simulated lorentz dist}
\end{figure}

For this work, we assume a wind duration of 50 seconds with a separation of 0.002 seconds between each shell launching. The Lorentz factor profile of the ejecta is a combination of multiple Fast Rise Exponential Decay (FRED) distributions, a distribution that can model smooth changes in the Lorentz factor profile over reasonable timescales (see Figure \ref{fig: simulated lorentz dist}; \citealt{2014ApJ...783...88H}). The outflow can be separated into `early' ejecta launched in the first 25 seconds and `late' ejecta launched for the remaining duration of the wind.

The early material has an average Lorentz factor of $\bar{\Gamma}_p\gtrsim150$ and represents the material responsible for the prompt emission. In this work, we do not consider, and remain agnostic to, the dissipation mechanisms responsible for producing the prompt emission. We utilize internal shocks in order to produce plateaus in the Lorentz factor profile of the late ejecta, but we do not require internal shocks to be the dominant dissipation mechanism of the prompt gamma-ray radiation. The later ejected material is launched with $\bar{\Gamma}_L\sim10$ and is responsible for the refreshed shocks (see Figure \ref{fig: ref-shock-schem}a). 

We find that the later ejecta undergoes internal shocks much earlier than the earlier ejected material, with respect to both time and distance from the central engine (in the source frame, see Figures \ref{fig: simulated lorentz dist} and \ref{fig: ref-shock-schem}b, c). 

This finding is supported if we use Equations \ref{eq: r coll is}, \ref{eq: t coll is}, and \ref{eq: t arriv} to estimate the radius (source frame) and observed collision time (observer frame) of two shells with $\Gamma_1$ and $\Gamma_2$ (where $\Gamma_2 > \Gamma_1$). If we consider two slow shells with Lorentz factors $\Gamma_1 = 10$ and $\Gamma_2 = 15$ are launched with a typical separation in time of $\delta t_e \sim 1$ sec, the radius and time of collision are $R_{coll} = 1.1 \times 10^{13}$ cm and $t_{a} = 0.8$ seconds, respectively. For two faster shells with Lorentz factors $\Gamma_1 = 100$ and $\Gamma_2 = 120$, launched with the same separation $\delta t_e$, we find $R_{coll} = 1.96\times10^{15}$ cm and $t_{a} = 2.3$ seconds. 

We let the ejecta propagate into a uniform density circumburst medium, i.e., $k=0$, with a particle density of $n = 1$ cm$^{-3}$. We find that the jet break time, i.e., when $\Gamma\theta_j = 1$, occurs at $t = 0.14$ days in our simulation (see Figure \ref{fig: ref-shock-schem}d), which is earlier than the observed jet break time of $\sim$0.5 days. However, the model light curve still shows a jet break that is similar to the observed jet break. This is most likely caused by the material which began emitting before jet break but was not finished emitting until some time after \citep{2011MNRAS.410.2016V}. 

As the outflow continues to propagate, the early ejecta continually slows down as it sweeps up material from the external medium. Eventually the early ejecta will decelerate until it is slow enough that the initially slow ejecta will catch up and collide with the material at the front of the jet (see Figure \ref{fig: ref-shock-schem}e and f).

\begin{figure}
	\centering
	\includegraphics[width=0.48\textwidth]{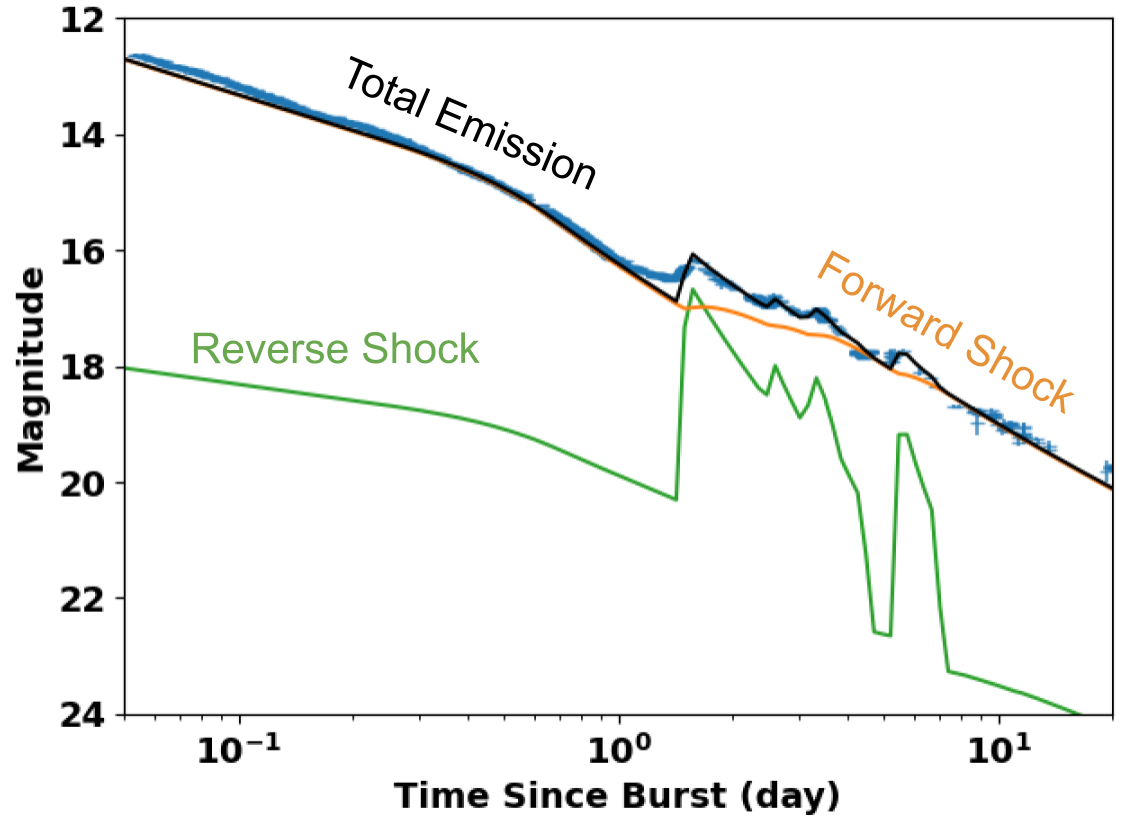}
	\caption{The optical afterglow light curve of GRB030329 (blue) and the total synthetic light curve (black) generated from the injected Lorentz profile (see Figure \ref{fig: simulated lorentz dist}) ejected into a uniform density medium. The contributions of the forward shock and reverse shocks are displayed in orange and green, respectively.}
	\label{fig: comb mag light curves}
\end{figure}

The simulated optical light curve from the collision exhibits jumps qualitatively consistent with the re-brightening events observed in the data at $\sim 1$ day (see Figure \ref{fig: comb mag light curves}). As argued at the end of Section \ref{sec: gamma disp}, we find that even if the late material is ejected with a wide spread of Lorentz factor values, internal shocks naturally narrow the Lorentz factor distribution of the material such that $\Delta\Gamma_{L}/\Gamma_{L}\lesssim0.1$ (see Figure \ref{fig: simulated lorentz dist}). 

We do not recover the slight flattening that occurs in the observed light curve just before the first jump. This flattening may be caused by a gradual increase in density leading up to the reverse shock, which would not be recovered by our simple model.

In Table \ref{tab: param values}, we display the micro- and macrophysical parameters used in our model to obtain the results displayed in Figure \ref{fig: comb mag light curves}. This combination of parameters does not uniquely describe the synthetic light curve. There is strong degeneracy within the parameter space. Changing the initial Lorentz factor profile or taking different values for the microphysical parameters will affect the characteristics of the refreshed shock emission (e.g., the shape, rise time, time of occurrence, amplitude) and can easily result in the emission becoming non-discernible above the afterglow emission continuum. This may explain why we do not often witness jumps in the optical afterglow light curve of GRBs.

\begin{figure}
	\centering
	\includegraphics[width=0.48\textwidth]{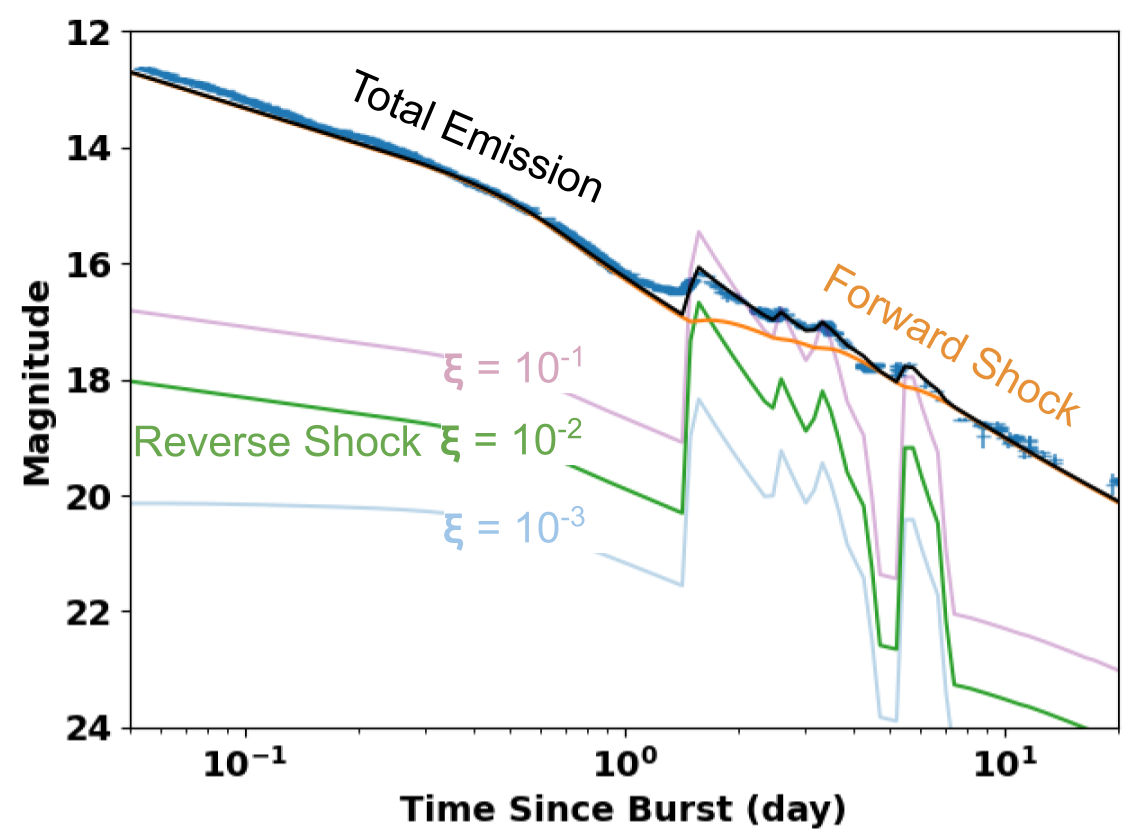}
	\includegraphics[width=0.48\textwidth]{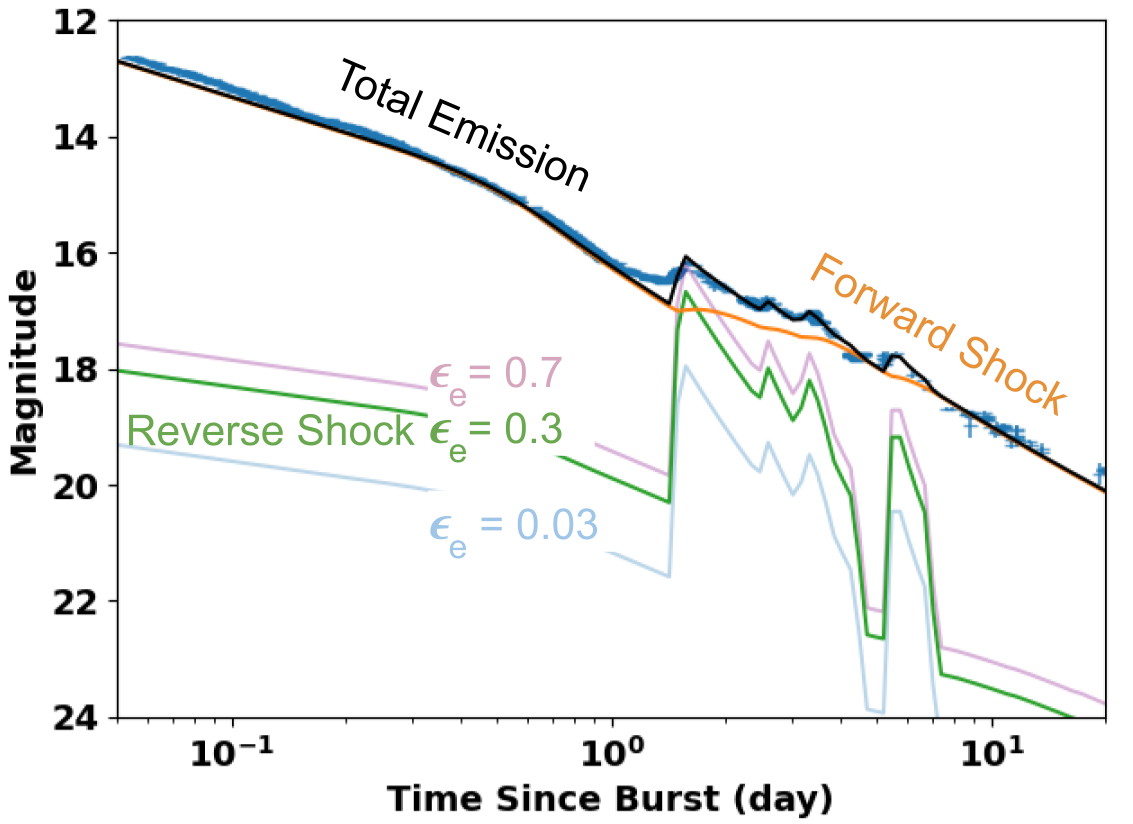}
	\caption{Same description as for Figure \ref{fig: comb mag light curves}, but additional reverse shock light curves have been included to visualize the influence of two microphysical parameters. All other values are the same as shown in Table \ref{tab: param values}. \textit{Top}: Reverse shock emission for different values of the fraction of electrons accelerated by the reverse shock, $\xi$, i.e., $\xi = 10^{-1}$ in pink and $\xi = 10^{-3}$ in light blue. As the fraction of electrons is increased, so too does the brightness of the reverse shock emission. \textit{Bottom}: Reverse shock emission for different values of the fraction of energy stored in the electron population, $\epsilon_{e}$, i.e., $\epsilon_{e} = 0.7$ in pink and $\epsilon_{e} = 0.03$ in light blue. As the fraction of energy stored in the electron population is increased, so too does the brightness of the reverse shock emission.}
	\label{fig: comb mag light curves micro comp}
\end{figure}

Some microphysical parameters strongly affect the brightness of the reverse shock emission compared to the forward shock emission and, for the parameter combination used in this simulation, must remain close to the values reported in Table \ref{tab: param values}. Other microphysical parameters have smaller effects on the light curve produced. Figure \ref{fig: comb mag light curves micro comp} shows how the reverse shock emission varies due to changing the value of $\xi$ and $\epsilon_e$. Determining which parameter combinations can lead to observable refreshed shock signatures requires a full parameter space exploration. This will be performed in future work.

\begin{table*}
    \centering
    \caption{Physical Parameters Used for Results Displayed in Figure \ref{fig: comb mag light curves}.}
    \label{tab: param values}
    \begin{tabular}{lll} 
        \hline
        Parameter & Variable & Value\\
        \hline
        Wind duration & $t_w$ & 50 sec \\
        Energy injection rate & $\dot{E}_{iso}$ & $4\times10^{51}$ erg s$^{-1}$ \\
        Jet opening angle & $\theta_j$ & 0.105 rad  \\
        Fraction of internal shock energy stored in electrons & $\epsilon_{e,ej}$ & 0.3 \\
        Fraction of internal shock energy stored in the magnetic field & $\epsilon_{B,ej}$ & 0.001 \\
        Fraction of electrons accelerated in internal shocks & $\xi_{ej}$ & $10^{-2}$ \\
        Power-law index of the electrons population distribution accelerated by internal shocks & $p_{ej}$ & 2.2 \\
        Fraction of external shock energy stored in electrons & $\epsilon_{e,cb}$ & 0.1 \\
        Fraction of external shock energy stored in the magnetic field & $\epsilon_{B,cb}$ & $10^{-4}$ \\
        Fraction of electrons accelerated in external shocks & $\xi_{cb}$ & $1$ \\
        Power-law index of the electrons population distribution accelerated by external shocks & $p_{cb}$ & 2.2\\
        \hline
    \end{tabular}
\end{table*}

\subsection{Are the Jumps the Signature of Forward or Reverse Shocks?} \label{sec: FS or RS} 

\citet{2003Natur.426..138G} assume that the jumps in the optical light curve are due to the forward shock emission. They further state that a possible prediction of the refreshed shock model is to expect flares to be observed at radio wavelengths as signatures of the reverse shock. \citet{2003Natur.426..154B} claim that the refreshed shock model cannot be viable because there is no observed reverse shock emission in the millimeter radio band. 

However, millimeter radio emission from a reverse shock may be subdominant to forward shock emission if the reverse shock accelerates a fraction of electrons smaller than unity, $\xi_{ej} < 1$ \citep{1998ApJ...496L...1R,2000ApJ...535L..33S}. 

\begin{figure}
	\centering
	\includegraphics[width=0.47\textwidth]{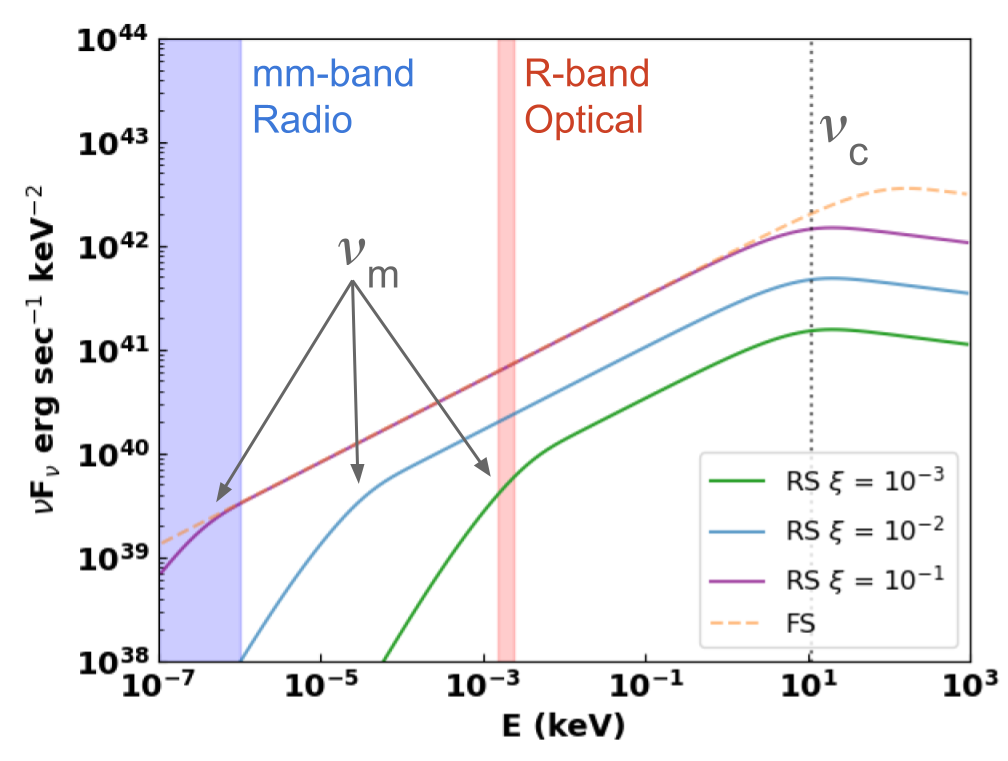}
	\caption{Synthetic spectra of the forward shock (dashed, orange line) and the reverse shocks for various values of $\xi_{ej}$, the fraction of electrons accelerated by the reverse shock (i.e., $\xi_{ej}=10^{-3}$ in green, $\xi_{ej}=10^{-2}$ in blue, and $\xi_{ej}=10^{-1}$ in purple) at the time of the first jump in the afterglow light curve, $\sim 1$ day. For increasing values of $\xi_{ej}$, $\nu_c$ (the synchrotron cooling frequency) remains the same for each curve, but $\nu_m$ (the frequency associated with the typical Lorentz factor of the electron population) moves to lower frequencies (see Equations \ref{eq: nu_c} and \ref{eq: nu_m}). The millimeter radio band is highlight by the blue region. The R-band optical regime is highlight by the red region.}
	\label{fig: spectrum at 1 day}
\end{figure}

Assuming the reverse shock emits synchrotron radiation, reducing $\xi_{ej}$ moves $\nu_m$ to higher frequencies and, in some cases, results in the break occurring at frequencies above the millimeter radio band. In Figure \ref{fig: spectrum at 1 day}, we display reverse shock spectra for various values of $\xi_{ej}$ and show that for $\xi_{ej} \lesssim 10^{-2}$ the reverse shock emission may be subdominant to the forward shock component at radio wavelengths. Furthermore, it has been shown in detailed hydrodynamic simulations that the density in the reverse shock material can be orders of magnitude higher than the average density of the ejecta \citep{2000A&A...358.1157D}. If the material crossed by the reverse shock radiates synchrotron emission while in the fast-cooling regime, increasing the density will further shift $\nu_m$ to higher frequencies (i.e., since $\nu_m \propto n^{1/2}$; see Equation \ref{eq: nu_m}). 

We find that the jumps in the optical afterglow of GRB030329 share similar rise times and peak shapes with jumps created from a combination of both forward and reverse shock emission, as opposed to purely forward shock emission (see Figure \ref{fig: comb mag light curves}). Forward shock emission tends to rise more slowly and smoothly, but is responsible for transitioning the outflow to a higher-energy Blandford-Mckee solution. The rise time of the forward shock is dependent on the energy ratio between the forward + reverse shock system and the injected material responsible for the refreshed shocks. Reverse shock emission displays quick rising flare events, possibly peaking above the forward shock for a brief time and then falling back below the afterglow continuum. The flux ratio between the forward and reverse shock emission is dependent on the micro- and macrophysical parameters assumed (see Figure \ref{fig: comb mag light curves micro comp}).

\section{Will the Emission of Late Ejecta Internal Shocks be Observable?} \label{sec: photosphere}


Synchrotron emission should be produced by late material after it has undergoes internal shocks. However, the internal shocks may occur below the photosphere of the material, i.e., the synchrotron emission may be unable to escape the emitting material, thus preventing the emission from being observable. We can estimate the location of the photosphere for a shell with Lorentz factor $\Gamma$ to be, 

\begin{align}
	R_{ph} \simeq \frac{\kappa \dot{E}_{kin}}{8\pi c^3 \Gamma^3} = 2.9\times 10^{15} \frac{\kappa_{0.2} \dot{E}_{iso,52}}{(1+\sigma)\Gamma_1^3} \text{ cm} \label{eq: phot radius}
\end{align}

where $\kappa$ is the opacity of the material ($\kappa_{0.2}$ is in units of 0.2 cm$^2$ g$^{-1}$) and $\Gamma$ is the Lorentz factor of the shell, and $\dot{E}_{iso}$ is the isotropic equivalent energy injection rate \citep{2013A&A...551A.124H}.

Approximating the Lorentz factor of the fast shell in the slow material as $\Gamma_f = 3\Gamma/2$, where $\Gamma$ is the Lorentz factor of a slower shell launched some time before the fast shell, we can approximate the radius of collision between the two shells Equation \ref{eq: r coll is} with the relation

\begin{align}
	R_{coll} \approx 4 c \delta t_e \Gamma^2
\end{align}
which, for a typical timescale of $\delta t_e = 1$ second, leads to the relation $R_{coll} \approx 1.2\times10^{13} \Gamma_1^2$ cm, where $\Gamma_1$ is the Lorentz factor of the slow shell in units of 10. The ratio of the internal shock radius to the photospheric radius can then be expressed as

\begin{align}
	\frac{R_{coll}}{R_{ph}} \approx  4\times10^{-3} \frac{( 1 + \sigma)}{\kappa_{0.2} \dot{E}_{iso,52}} \Gamma_1^5 \text{ cm}
\end{align}

Taking reasonable values for the average Lorentz factor of the late ejecta to be $\Gamma = 10$, the magnetization to be $\sigma = 0.1$ (the upper limit of $\sigma$ that allows for strong reverse shocks; \citealt{2015SSRv..191..519S}), and taking a reasonable value for the isotropic equivalent injected energy, $\dot{E}_{iso} = 10^{52}$ erg s$^{-1}$, the photospheric radius of the slow, late ejecta will be larger than its own internal collision radius, implying that the internal shock emission of the late ejecta will not be observable because it occurs below its own photosphere. 

The luminosity of thermal emission coming from a jet photosphere can be approximated using Equation \ref{eq: phot radius} and Equation 8 from \citet{2013A&A...551A.124H}. Taking typical values of the jet opening radius to be $\ell=10^6$ cm, $\sigma = 0.1$, $\kappa = 0.2$ cm$^2$ g$^{-1}$, and $\theta_j = 0.1$ rad, the photospheric luminosity can be approximated to be
\begin{align}
	L_{TH} \approx 6 \times 10^{46} \dot{E}_{iso,52}^{1/3} \Gamma_1^{8/3} \text{ erg/s}
\end{align}
thus we do not expect observable thermal emission from the photosphere of the slow ejecta (however, see \citet{2016MNRAS.457L.108B} for a model that explains X-ray flares using the photospheric emission of slowly moving outflow material). 


\section{Conclusion} \label{sec: conclusion}

We summarize our findings here:

\begin{enumerate}
	\item Slow material launched by the central object slightly ($\sim$a few seconds) after the material responsible for the main prompt emission may be responsible for quickly rising energy injection events observed in GRB optical afterglow light curves. A long-lived central engine is not necessary for late-time behavior in the afterglow light curve \cite{2016MNRAS.457L.108B}.
	\item A quickly rising energy injection period requires the injected material to have a narrowly distributed Lorentz distribution, which is naturally produced via internal shocks.
	\item If material is launched with a low Lorentz factor, internal shocks will occur close to the central engine and at early times, indicating that the magnetization $\sigma$ at these small radii and early times must be sufficiently low to allow for shocks to form (i.e., $\sigma < 0.1$).
	\item The radius at which the slow material undergoes internal shocks will most often be below its own photosphere, so it should not be expected to observe bright prompt emission coming from the slow material.
\end{enumerate}

Although there are many GRBs which display flaring activities or energy injection periods in their optical light curves \citep{2005A&A...439L..35K,2007ApJ...671.1921F,2012ApJ...758...27L,2013ApJ...774...13L,2013ApJ...774....2S,2014ApJ...788...30S}, observations of late-time, quick-rising energy injection events are rarely seen in GRB optical afterglows. This may imply that slowly moving material ejected by the central engine is not common in GRBs or that the physical parameter combination are such that the variations in optical flux are not significant or observable. 

We find that the shape, rise time, time of occurrence, and amplitude of the refreshed shock jumps are highly influenced by the values of the physical parameters in the outflow material and by the initial Lorentz factor profile of the ejecta. Slightly changing the values of the microphysical parameters or Lorentz profile can easily result in the emission from the shocks becoming non-discernible above the afterglow emission continuum. So, perhaps it should not be expected to often observe similar jumps in optical afterglow light curves of GRBs because they require such specific conditions. Barring any physical reasoning, the lack of observed afterglow jumps may simply be due to the small sample of highly time-resolved optical light curves observed during the first few days following the GRB prompt emission. 

Our model assumes 1D planar shells of cold material, so hydrodynamic timescales are under estimated during shell collisions. However, our estimates of the relevant hydrodynamic timescale, such as the reverse shock shell crossing time, show that the rise times in our model are still accurate to within a factor of unity (see Section \ref{sec: rise times}).

One alternative explanation to produce the $\Delta t / t < 1$ timescales observed for the jumps could also be to assume that the emission is not distributed isotropically in the shock rest frame \citep{2011MNRAS.410.2422B}. This still requires some energy injection mechanisms to produce the jumps but does not require the jumps to occur after jet break. Synchrotron radiation is oriented relative to the magnetic field in the emitting region and is, therefore, anisotropic, but the orientation of magnetic fields in GRB outflows are likely highly variable.

In this work, we have confirmed that the refreshed shocks model proposed by \citet{2003Natur.426..138G} can be used to explain the origin of the multiple rebrightening events in the optical afterglow of GRB030329. The model requires the Lorentz factor distribution of the refreshed shock material to be relatively narrow to produce the fast rise-times of the jumps, a condition that naturally occurs if the material undergoes internal shocks. The presence of internal shocks implies that the magnetization of the material must be sufficiently low to allow for efficient acceleration of charged particles via internal shocks (e.g., $\sigma < 0.1$).

\section*{Acknowledgements}

MM would like to thank Cl\'ement Pellouin for helpful discussions. M.M. and S.G. thank the Chateaubriand Fellowship Program supported by the Embassy of France in Washington for funding part of this work in enabling M.M.'s stay at Institut d'Astrophysique de Paris for a period of 9 months. PB was supported by a grant (no. 2020747) from the United States - Israel Binational Science Foundation (BSF), Jerusalem, Israel. F.D. and R.M. acknowledge financial support from the Centre National d’\'Etudes Spatiales (CNES). S.G. acknowledges financial support through the Cycle-18 NASA Neil Gehrels Swift Observatory Guest Investigator program.

\section*{Data Availability}

The model data underlying this article will be shared on reasonable request to the corresponding author. The data were derived from previously published works \citep{2003Natur.423..843U,2003AstL...29..573B,2003ApJ...599..394M,2004AJ....127..252B,2004ApJ...606..381L}.


\bibliographystyle{mnras}
\bibliography{bibliograpy-list}

\appendix

\section{Schematic of The Refreshed Shock Model}
\begin{figure*}
	\centering
	\includegraphics[width=0.49\linewidth]{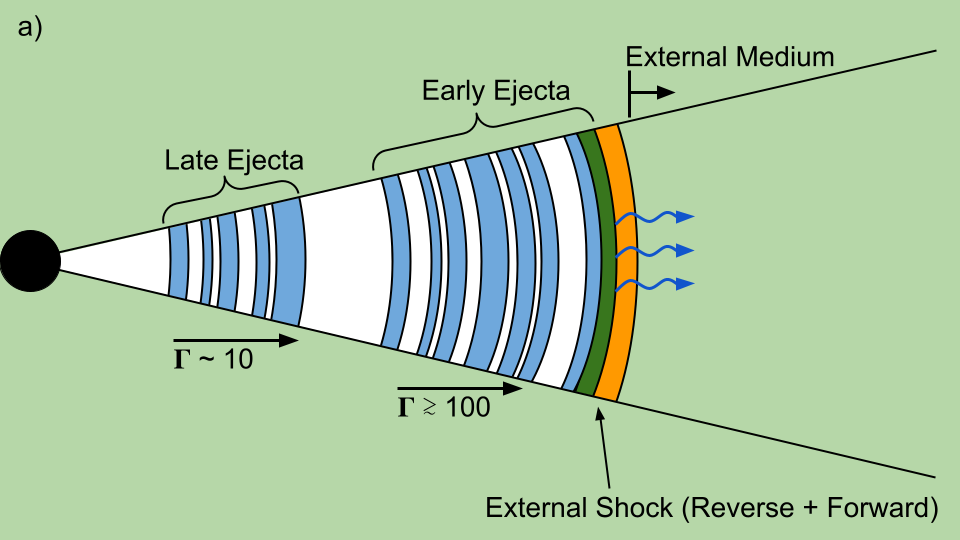}
	\includegraphics[width=0.49\linewidth]{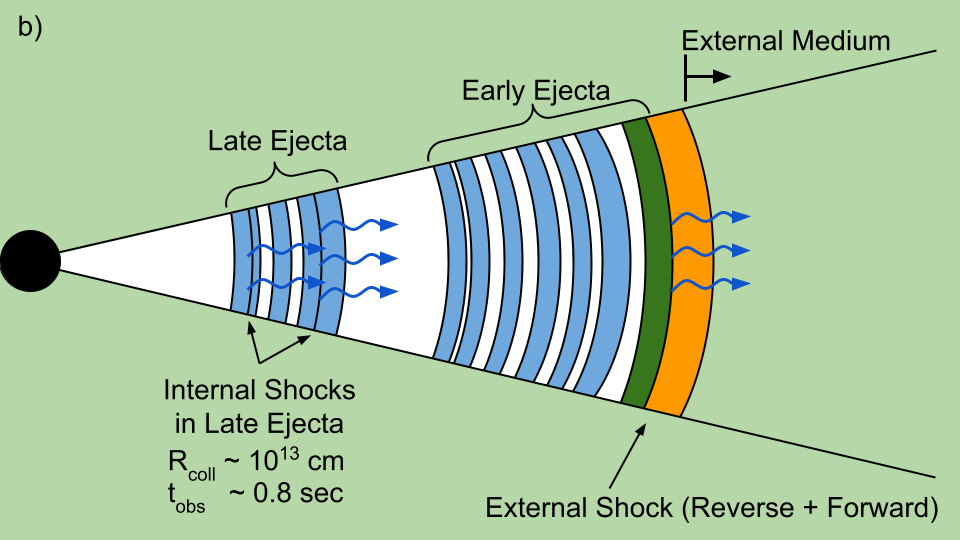}
	\includegraphics[width=0.49\linewidth]{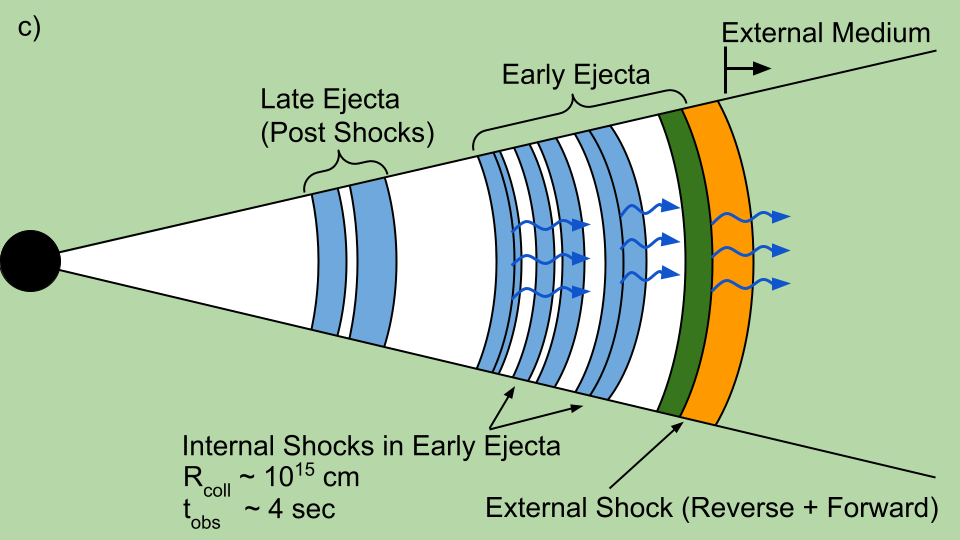}
	\includegraphics[width=0.49\linewidth]{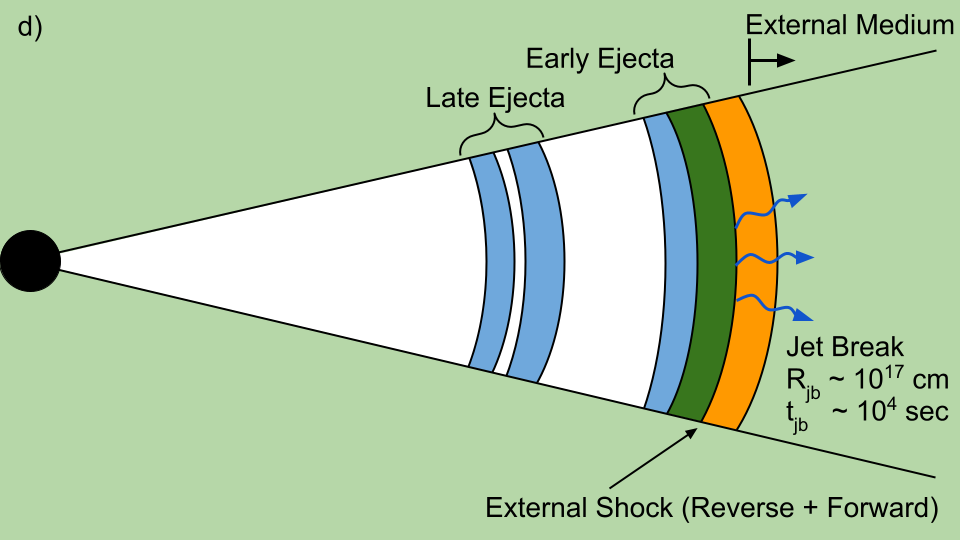}
	\includegraphics[width=0.49\linewidth]{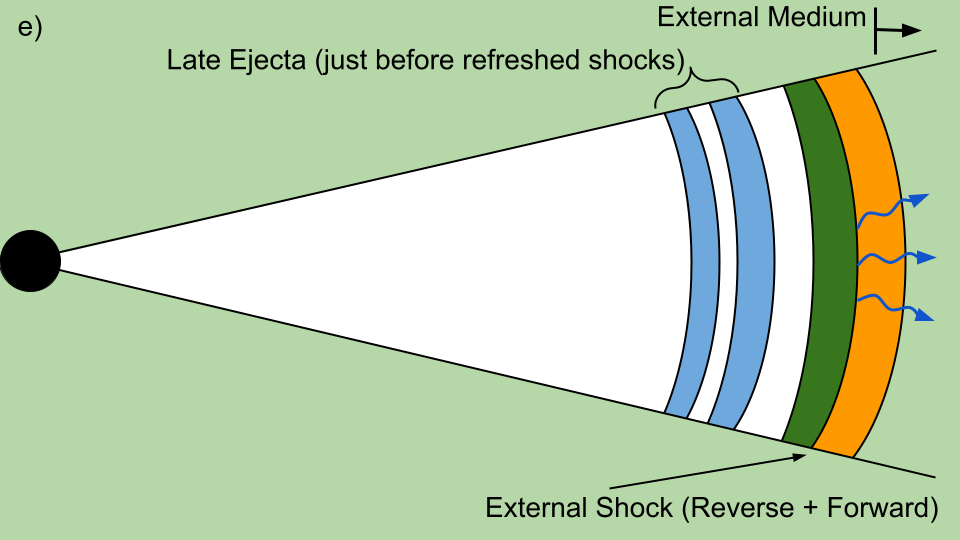}
	\includegraphics[width=0.49\linewidth]{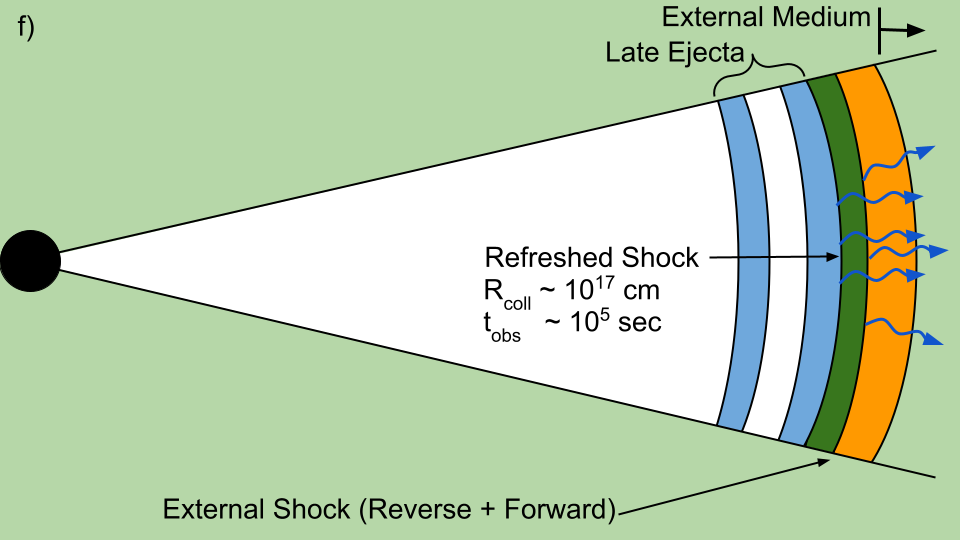}
	\caption{a) The early ejecta is launched by the central engine with a bulk Lorentz factor of $\Gamma \geq 100$. It quickly encounters the circumburst medium and forms a forward shock propagating through the external medium. As the outflow sweeps up circumburst material, it decelerates and emits radiation, i.e., afterglow emission. A reverse shock is formed as material in the outflow catches up and collides with the decelerating jet front. Some time later, material is launched from the central engine with bulk Lorentz factors $\Gamma\sim10$. This late ejecta travels in the wake of the earlier material so it does not encounter any circumburst medium and does not decelerate. b) Due to its low speeds, the late ejecta quickly undergoes internal shocks. c) The early ejecta will continue to collide with the front of the jet via a reverse shock and will eventually undergo internal shocks of its own, producing the prompt emission radiation. d) Some time later the jet will experience a jet break (e.g., when $\Gamma\theta_j \lesssim 1$). e) Eventually all of the earlier ejecta will be crossed by the reverse shock and the only emission will be that of the decelerating front of the jet as more circumburst material is continually swept up. f) The jet front will eventually decelerate so much that the late ejecta will catch up and ram the front of the jet, i.e., refreshed shocks. The refreshed shocks can be witnessed as jumps in the optical afterglow light curve. Approximate radii (source frame) and time of observation (observer frame) are displayed for critical points in the dynamical evolution.}
	\label{fig: ref-shock-schem}
\end{figure*}

A schematic of the refreshed shock model is shown in Figure \ref{fig: ref-shock-schem}. 

\bsp	
\label{lastpage}
\end{document}